\documentclass[%
reprint,
superscriptaddress,
amsmath,
amssymb,
aps
]{revtex4-2}

\usepackage{graphicx}
\usepackage{dcolumn}
\usepackage{bm}

\usepackage{mathptmx}
\graphicspath{{images/}}
\usepackage{textgreek}
\usepackage{placeins}
\usepackage{color}
\usepackage{xcolor}
\usepackage{soul}

\begin{document}

\title{Two-channel anomalous Hall effect in  SrRuO\textsubscript{3}}

\author{Graham Kimbell}
\affiliation{Department of Materials Science \& Metallurgy, University of Cambridge, CB3 0FS, United Kingdom}
\author{Paul M. Sass}
\affiliation{Department of Physics \& Astronomy, Rutgers University, Piscataway, NJ 08854, USA}
\author{Bart Woltjes}
\affiliation{Department of Materials Science \& Metallurgy, University of Cambridge, CB3 0FS, United Kingdom}
\author{Eun Kyo Ko}
\affiliation{Center for Correlated Electron Systems, Institute for Basic Science (IBS), Seoul 08826, Republic of Korea}
\affiliation{Department of Physics \& Astronomy, Seoul National University, Seoul 08826, Republic of Korea}
\author{Tae Won Noh}
\affiliation{Center for Correlated Electron Systems, Institute for Basic Science (IBS), Seoul 08826, Republic of Korea}
\affiliation{Department of Physics \& Astronomy, Seoul National University, Seoul 08826, Republic of Korea}
\author{Weida Wu}
\affiliation{Department of Physics \& Astronomy, Rutgers University, Piscataway, NJ 08854, USA}
\author{Jason W. A. Robinson}
\email[]{jjr33@cam.ac.uk}
\affiliation{Department of Materials Science \& Metallurgy, University of Cambridge, CB3 0FS, United Kingdom}

\date{\today}

\begin{abstract}
The Hall effect in SrRuO\textsubscript{3} thin-films near the thickness limit for ferromagnetism shows an extra peak in addition to the ordinary and anomalous Hall effects. This extra peak has been attributed to a topological Hall effect due to two-dimensional skyrmions in the film around the coercive field; however, the sign of the anomalous Hall effect in SrRuO\textsubscript{3} can change as a function of saturation magnetization. Here we report Hall peaks in SrRuO\textsubscript{3} in which volumetric magnetometry measurements and magnetic force microscopy indicate that the peaks result from the superposition of two anomalous Hall channels with opposite sign. These channels likely form due to thickness variations in SrRuO\textsubscript{3}, creating two spatially separated magnetic regions with different saturation magnetizations and coercive fields. The results are central to the development of strongly correlated materials for spintronics.
\end{abstract}

\maketitle

\section{Introduction}

\begin{figure}[htbp]
\includegraphics[width=223pt]{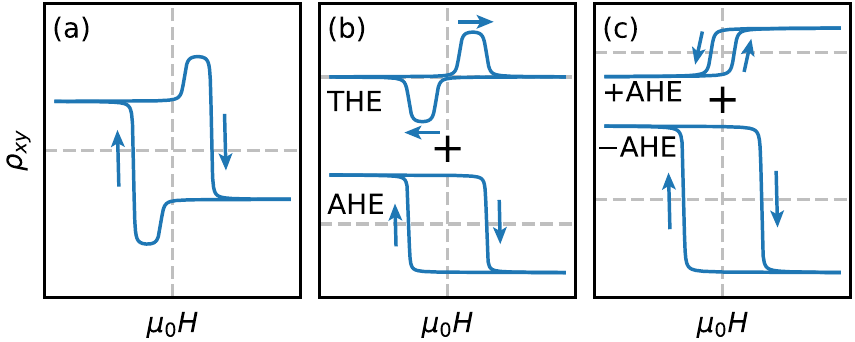}
\caption{(a) Illustrations of the peak structure in the Hall effect describing how the peaks may arise due to (b) the superposition of the AHE and THE, or (c) two AHE channels with different signs and coercive fields.}
\label{fig:Hall effect models}
\end{figure}

In the Hall effect a transverse electric field ($E_x$) is generated under an applied longitudinal current density ($J_y$). In conventional tensor notation, $E_x = \rho_{xy} J_y$ where $\rho_{xy}$ is the Hall resistivity. In a ferromagnet, $\rho_{xy} = R_0 H_z + R_s M_z$, the ordinary Hall effect (OHE) coefficient ($R_0$) is proportional to the out-of-plane applied magnetic field ($H_z$), is caused by the Lorentz force, and its sign follows the sign of the charge carriers; the anomalous Hall effect (AHE) coefficient ($R_s$) is proportional to the out-of-plane component of magnetization ($M_z$). Recently, there have been reports of an additional peak in the Hall resistivity (illustrated in Fig.~\ref{fig:Hall effect models}(a)) in ultrathin films of ferromagnetic SrRuO\textsubscript{3} \cite{matsuno_interface-driven_2016, ohuchi_electric-field_2018,meng_observation_2019,sohn_emergence_2018,gu_interfacial_2019, qin_emergence_2019, sohn_hump-like_2020,zhang_robust_2020, wang_ferroelectrically_2018,wang_spin_2019,li_reversible_2020,sohn_hump-like_2020,gu_interfacial_2020,groenendijk_berry_2018,kan_alternative_2018,malsch_correlating_2019,ren_nonvolatile_2019,wang_controllable_2019,wu_berry_2019,ziese_unconventional_2019,wu_thickness_2020,kan_field-sweep-rate_2020}. This additional peak is sometimes attributed to a spin texture with net spin chirality \cite{matsuno_interface-driven_2016, ohuchi_electric-field_2018,meng_observation_2019,sohn_emergence_2018,gu_interfacial_2019, qin_emergence_2019, sohn_hump-like_2020,zhang_robust_2020, wang_ferroelectrically_2018,wang_spin_2019,li_reversible_2020,sohn_hump-like_2020}, which results in electrons experiencing an effective magnetic field from the accumulation of a Berry phase in real space. This additional contribution to the Hall effect is called the topological Hall effect (THE) \cite{bruno_topological_2004} (Fig.~\ref{fig:Hall effect models}(b)). A possible source of chirality is skyrmions in the SrRuO\textsubscript{3}, which are topologically protected magnetic textures. Skyrmions can be stabilized through a competition between the ferromagnetic exchange and the Dzyaloshinkii-Moria interaction (DMI) resulting from the combination of spin-orbit coupling and broken inversion symmetry. The source of spin-orbit coupling is either extrinsic from a heavy-metal oxide layer such as SrIrO\textsubscript{3} \cite{matsuno_interface-driven_2016, ohuchi_electric-field_2018, meng_observation_2019}, or intrinsic to the SrRuO\textsubscript{3} \cite{sohn_emergence_2018,gu_interfacial_2019, qin_emergence_2019, sohn_hump-like_2020,zhang_robust_2020}. The broken inversion symmetry arises from the interfaces between SrRuO\textsubscript{3} and adjacent layers. It is also possible that the DMI is sufficient that domain walls around bubble domains have net chirality and therefore give a THE contribution \cite{jiang_crossover_2019}.

Alternatively, the peak in the Hall effect could arise due to the superposition of two AHEs with different signs \cite{gerber_interpretation_2018,groenendijk_berry_2018,kan_alternative_2018,wu_artificial_2018,malsch_correlating_2019,ren_nonvolatile_2019,wang_controllable_2019,wu_berry_2019,ziese_unconventional_2019,wu_thickness_2020,kan_field-sweep-rate_2020} (Fig.~\ref{fig:Hall effect models}(c)). The AHE in SrRuO\textsubscript{3} has a non-monotonic temperature and resistivity dependence which cannot be explained by scattering mechanisms alone \cite{berger_side-jump_1970,smit_spontaneous_1955,smit_spontaneous_1958}, instead, it is thought to also depend on the Berry curvature in $k$-space  \cite{onoda_topological_2002,fang_anomalous_2003,jungwirth_anomalous_2002,haham_scaling_2011}. This is an intrinsic mechanism for the Hall effect and should not be confused with the extrinsic accumulation of Berry phase in real space by skyrmions. The anomalous Hall coefficient in SrRuO\textsubscript{3} is sensitively dependent on the band structure and magnetization \cite{fang_anomalous_2003, onoda_topological_2002}, and is thus dependent on temperature ($T$) and thickness, as well as strain, disorder and stoichiometry. Crucially, the AHE in SrRuO\textsubscript{3} undergoes a sign-change as a function of saturation magnetization \cite{mathieu_scaling_2004,klein_extraordinary_2000,izumi_magnetotransport_1997}. Therefore, if the SrRuO\textsubscript{3} is inhomogeneous, consisting of two magnetic regions with different coercive fields and saturation magnetizations giving opposite sign AHEs, a peak will necessarily appear in the Hall effect.

In this paper, we show that the additional peak in the SrRuO\textsubscript{3} Hall effect is likely caused by the superposition of two opposite sign AHEs from two magnetic regions with different saturation magnetizations. We argue that these two regions arise from single unit-cell thickness variations in SrRuO\textsubscript{3} films that are close to the thickness limit for ferromagnetism. SrRuO\textsubscript{3} films between 4 and 5 pseudocubic unit cells (UC) thick reproducibly display additional peaks in Hall effect measurements. Volumetric magnetometry shows two magnetic transition temperatures and two switching fields consistent with the Hall effect, and magnetic force microscopy (MFM) confirms these two regions are spatially separated and magnetically non-uniform.

\section{Methods}

SrRuO\textsubscript{3} films are grown by pulsed laser deposition on TiO\textsubscript{2}-terminated (001)-oriented SrTiO\textsubscript{3}. The SrTiO\textsubscript{3} is held at 600$^{\circ}$C during deposition under a flow of 100 mTorr O\textsubscript{2}, then cooled in 400~mTorr O\textsubscript{2} at $-5^{\circ}$C~min$^{-1}$. Laser pulses of 10~Hz with an energy density $\approx$ 2.5~J~cm$^{-2}$ give a growth rate of one UC ($c = 3.95$~\AA) per 10 seconds. SrRuO\textsubscript{3} grows initially layer-by-layer then transitions to step-flow-growth after several UCs \cite{choi_growth_2001}. In the first layer, the substrate changes from B-site (TiO\textsubscript{2}) to A-site (SrO) termination \cite{rijnders_enhanced_2004}. This can be considered as half-integer UC thicknesses of SrRuO\textsubscript{3}, or as the first half UC being a continuation of the SrTiO\textsubscript{3} substrate; the latter definition is used here, as depicted in Fig.~\ref{fig:AFM1}(c). The growth rate (and substrate terrace width) is high enough to reduce growth instabilities caused by the strain in SrRuO\textsubscript{3} \cite{esteve_step_2011}, but low enough to allow for diffusion of adatoms across terraces \cite{rao_growth_1997}. TiO\textsubscript{2}-terminated SrTiO\textsubscript{3} ensures uniform nucleation and growth of SrRuO\textsubscript{3} \cite{bachelet_self-assembly_2009}.

The structural quality of films is assessed by X-ray diffraction (XRD), X-ray reflectivity (XRR), and atomic force microscopy (AFM). The growth rate is calibrated by fitting fringes in XRR and XRD for thicker samples (see Supplementary Information).

Volumetric magnetic properties are investigated using a superconducting quantum interference device with films loaded out-of-plane with respect to the applied magnetic field. To identify background signals, measurements were taken of both films, as-received substrates, and substrates which underwent the same growth and cleaning conditions as the film without any material deposited (see Supplementary Information).

Transport measurements are made using unpatterned van der Pauw geometries. The transverse resistance data is anti-symmetrised to separate the Hall effect from the longitudinal component of resistance (see Supplementary Information).

Micromagnetic surface measurements are carried out using a cryogenic MFM with piezoresistive cantilevers, recorded in constant height mode with the scanning plane 100~nm above the SrRuO\textsubscript{3} film surface. For further details see \cite{sass_magnetic_2019}.

\section{Results \& discussion}

SrRuO\textsubscript{3} films are epitaxial and fully strained to the SrTiO\textsubscript{3} substrate (see Supplementary Information). Fig.~\ref{fig:AFM1}(a) shows an AFM image of a 4.5~UC thick film with an atomically flat surface and a step-and-terrace structure. A line profile (Fig.~\ref{fig:AFM1}(b)) shows the steps are one UC in height ($c=3.95$~\AA), indicating a single surface termination in the SrRuO\textsubscript{3}. Strain-induced growth instabilities in SrRuO\textsubscript{3} usually manifest as curvature at step edges.

\begin{figure}[htbp]
\includegraphics[width=246pt]{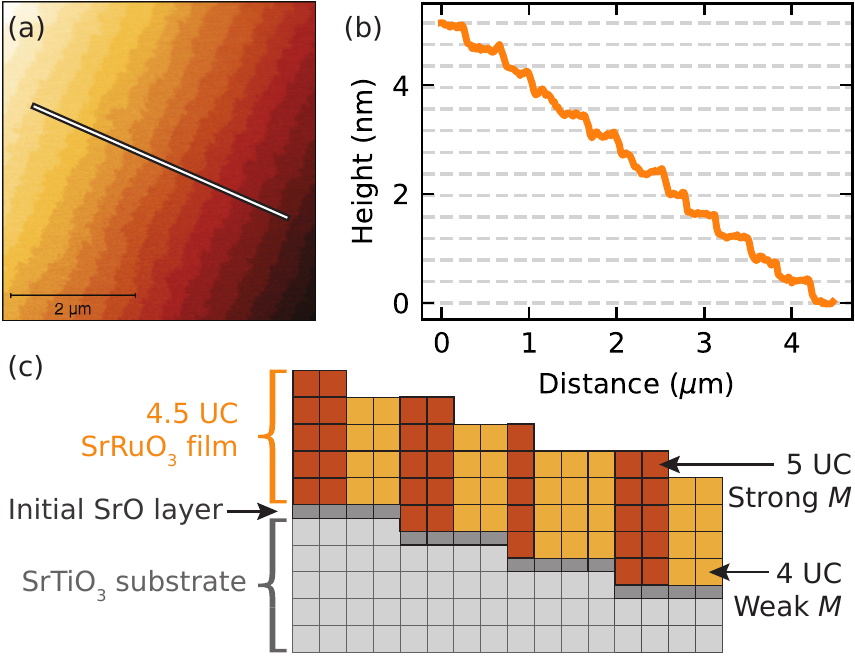}
\caption{(a) Plane-leveled AFM image of a 4.5~UC thick SrRuO\textsubscript{3} film, showing a step and terrace structure typical of step-flow growth. (b) The corresponding height profile along the line drawn in (a), the horizontal dotted lines are separations of one UC ($c=3.95$~nm, as measured in XRD on thicker films). (c) Illustration of thickness variations in the film.}
\label{fig:AFM1}
\end{figure}

\begin{figure}[!htbp]
\includegraphics[width=246pt]{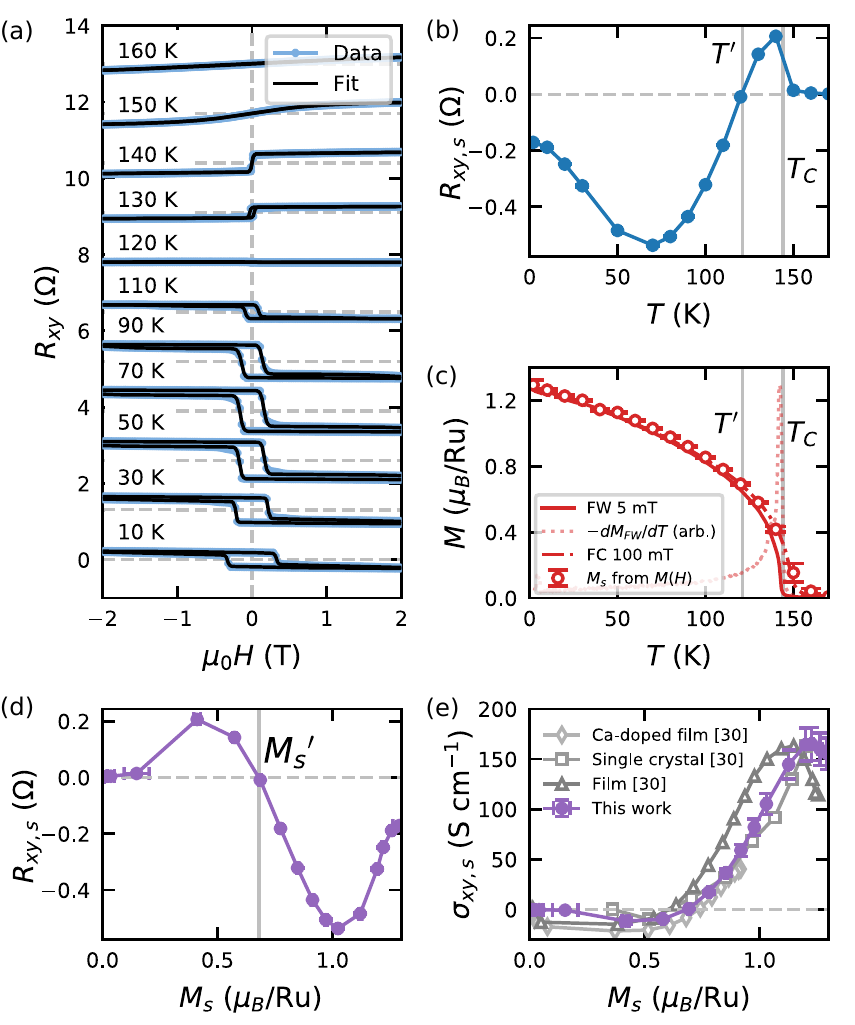}
\centering
\caption{Transport and magnetometry of a 10~nm thick film. (a) $R_{xy}(H)$ at different $T$. The OHE is removed by a linear fitting at high field (5-6~T). (b) The spontaneous anomalous Hall effect extracted from (a) as a function of temperature. This is non-monotonic with a sign change at $T' \approx 121$~K. (c) Magnetization as a function of temperature. Field cool and field warming curves have substrate curves subtracted and are shifted vertically so $M(170)=0$, the dotted curve shows the derivative of the 5~mT field warm curve with arbitrary units to highlight the magnetic transition. Points from $M(H)$ curves are spontaneous magnetization taken as average moment under 1~T for saturated quadrants. There is no magnetic transition evident at the AHE sign-change. (d) Spontaneous anomalous Hall resistivity of the SrRuO\textsubscript{3} as a function of spontaneous magnetization. The anomalous Hall effect is $+$ve for small $M_s$, and $-$ve for large $M_s$, switching at $M_s$$' \approx 0.7$~$\mu_B$/Ru. (e) Spontaneous anomalous Hall conductivity as a function of spontaneous magnetization, compared to previous literature values \cite{fang_anomalous_2003}.
}
\label{fig:RH1}
\end{figure}

Electronic transport of a 10~nm thick SrRuO\textsubscript{3} film (Figs.~\ref{fig:RH1}(a) and \ref{fig:RH1}(b)) show a spontaneous Hall resistance ($R_{xy,s}$) attributed to the AHE, which changes sign from positive ($+$ve) at high $T$ to negative ($-$ve) at low $T$. This unusual $T-$dependence has been observed previously \cite{klein_extraordinary_2000,fang_anomalous_2003} and is related to the intrinsic nature of the AHE in SrRuO\textsubscript{3}. We note that the apparent two magnetic switches seen in this case are likely caused by the rotation of the easy axis in SrRuO\textsubscript{3} away from the out-of-plane direction with increasing thickness, and not by two magnetic regions \cite{schultz_magnetic_2009}. The spontaneous volume magnetization ($M_s$) of a film is used as a scaling parameter for the AHE, so the magnetization of the 10~nm film is measured versus $T$ (Fig.~\ref{fig:RH1}(c)), showing only one transition at the Curie temperature ($T_C \approx 144$~K). The spontaneous Hall resistivity ($\rho_{xy,s}$) and conductivity (estimated as $\sigma_{xy,s} \approx -\rho_{xy,s}/\rho_{xx}^2$ where $\rho_{xx}$ is the longitudinal resistivity) are plotted in Figs.~\ref{fig:RH1}(d) and \ref{fig:RH1}(e). The Hall conductivity as a function of magnetization agrees well with literature values \cite{fang_anomalous_2003} despite differences in material properties. The anomalous Hall resistivity switches from $+$ve at low $M_s$, to  $-$ve at high $M_s$.

\begin{figure}[!htbp]
\includegraphics[width=246pt]{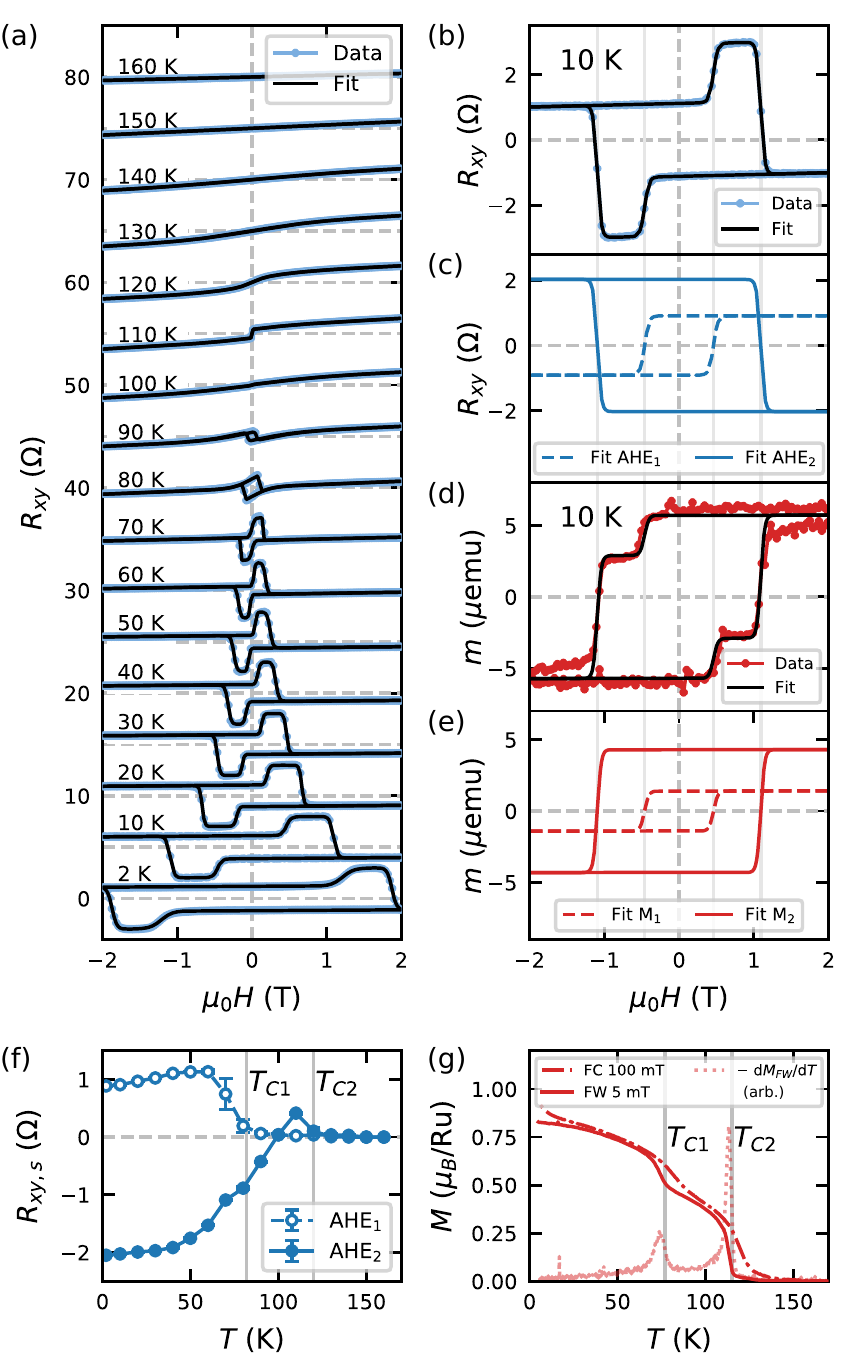}
\centering
\caption{Transport and magnetometry of a 4.5~UC thick SrRuO\textsubscript{3} film. (a) $R_{xy}$($H$) at different $T$. The data (blue points) are fitted to a function of two sigmoids as an approximation of two AHEs (black line). (b) $R_{xy}(H)$ of the film at 10~K, the black line is the two AHE fit. (c) The two fit components from from (b). (d) Moment ($m$) versus $H$ of the film at 10~K, showing two switches at the same coercive fields as in the Hall resistance. For the fit, the widths and coercive fields from (b,c) are fixed, and the scalings are fit parameters. (e) The two fit components from (d). (f) $T-$dependence of the spontaneous anomalous Hall resistance of the two channels, defined as the mean and standard deviation of anomalous Hall resistance in the 0$-$100~mT range of the saturated quadrant of an individual fit component. (g) $T-$dependence of the magnetization in field cooling (FC) in 100~mT, and field warming (FW) in 5~mT after saturating at 7~T. Data have substrate curves subtracted and are shifted vertically so $M(170)=0$. The dotted curve shows the derivative of the 5~mT field warm curve with arbitrary units to highlight the two magnetic transitions.
\newline
\newline
\newline
\newline
}
\label{fig:RH2}
\end{figure}

Hall measurements of a 4.5~UC thick film shows additional peaks in $R_{xy}$, similar to those attributed to a THE. Fig.~\ref{fig:RH2}(a) shows the $T-$dependence of $R_{xy}$($H$), with peaks below 90~K; these data are fitted assuming the two-channel AHE model, and Fig.~\ref{fig:RH2}(b) is the same data at only 10~K. The data is well reproduced by these two AHEs (Fig.~\ref{fig:RH2}(c)). The spontaneous anomalous Hall resistances of the two channels extracted from the fit is shown as a function of $T$ in Fig.~\ref{fig:RH2}(f). The $T-$dependence of the two AHE channels can be compared to that of the thicker film (Fig.~\ref{fig:RH1}(b)): a similar $T-$dependence is seen but suppressed to lower $T$ slightly in AHE\textsubscript{2}, and suppressed significantly in AHE\textsubscript{1}.

An $M(H)$ loop at 10~K is shown in Fig.~\ref{fig:RH2}(d) (for full temperature range see Supplementary Information). This is fitted using the same function as for the anomalous Hall effect, where only the scaling of the two components are varied as free parameters; these two components are shown in Fig.~\ref{fig:RH2}(e). There are two clear switches in the magnetometry which correspond to the two switches in Hall effect, and the smaller magnetic switch gives a $+$ve AHE component, whilst the larger switch gives a $-$ve AHE, as expected.

The magnetization versus temperature is shown in Fig.~\ref{fig:RH2}(g); there appears to be two separate magnetic transitions which is most obvious in the remanence curve (saturating with 7~T and warming in a 5~mT field, solid curve) and its derivative (dotted curve, arbitrary units and flipped for clarity), which show two transitions at $T_{C1} \approx 77$~K and $T_{C2} \approx 115$~K, labeled on the graph with solid gray lines. These two transition temperatures are also plotted in Fig.~\ref{fig:RH2}(f) and agree with the appearance of the spontaneous components of each AHE channel, so we attribute these transitions to the Curie temperatures for the two magnetic regions. It is likely that the 4~UC regions correspond to the lower $T_C$, lower $H_c$ and $+$ve AHE, whilst the 5~UC regions correspond to the higher $T_C$, higher $H_c$ and $-$ve AHE (at low $T$).

\medskip

A SrRuO\textsubscript{3} film prepared in a different system and measured sooner after deposition also shows two switches in $R_{xy}(H)$ (Fig.~\ref{fig:SNU}(a)), $R_{xy}(H)$ (Fig.~\ref{fig:SNU}(f)), $M(H)$, and $M(T)$ (see Supplementary Information). However, the Hall effect transitions from a peak-like structure at high $T$ (one low $M_s$ $+$ve AHE region, one high $M_s$ $-$ve AHE region, Figs.~\ref{fig:SNU}(b) and \ref{fig:SNU}(c)) to a step-like structure at low $T$ (two high $M_s$ $-$ve AHE regions, Figs.~\ref{fig:SNU}(d) and \ref{fig:SNU}(e)). This step-like structure is consistent with two AHEs. To explain this in terms of the THE, the OHE or spin polarisation of the SrRuO\textsubscript{3} would change sign at $\approx 53$~K. An OHE sign change at this $T$ is not observed here (Fig.~\ref{fig:SNU}(g)), but we note that though generally assumed to be constant below TC, the temperature dependent spin-polarisation of SrRuO3 is not well characterized theoretically or experimentally. An alternative explanation might be that the THE peak is shifting to higher fields above the coercive field of SrRuO\textsubscript{3}.This is, however, unlikely since an intermediate peak plus step shape is expected in this case.

\begin{figure}[!htbp]
\includegraphics[width=246pt]{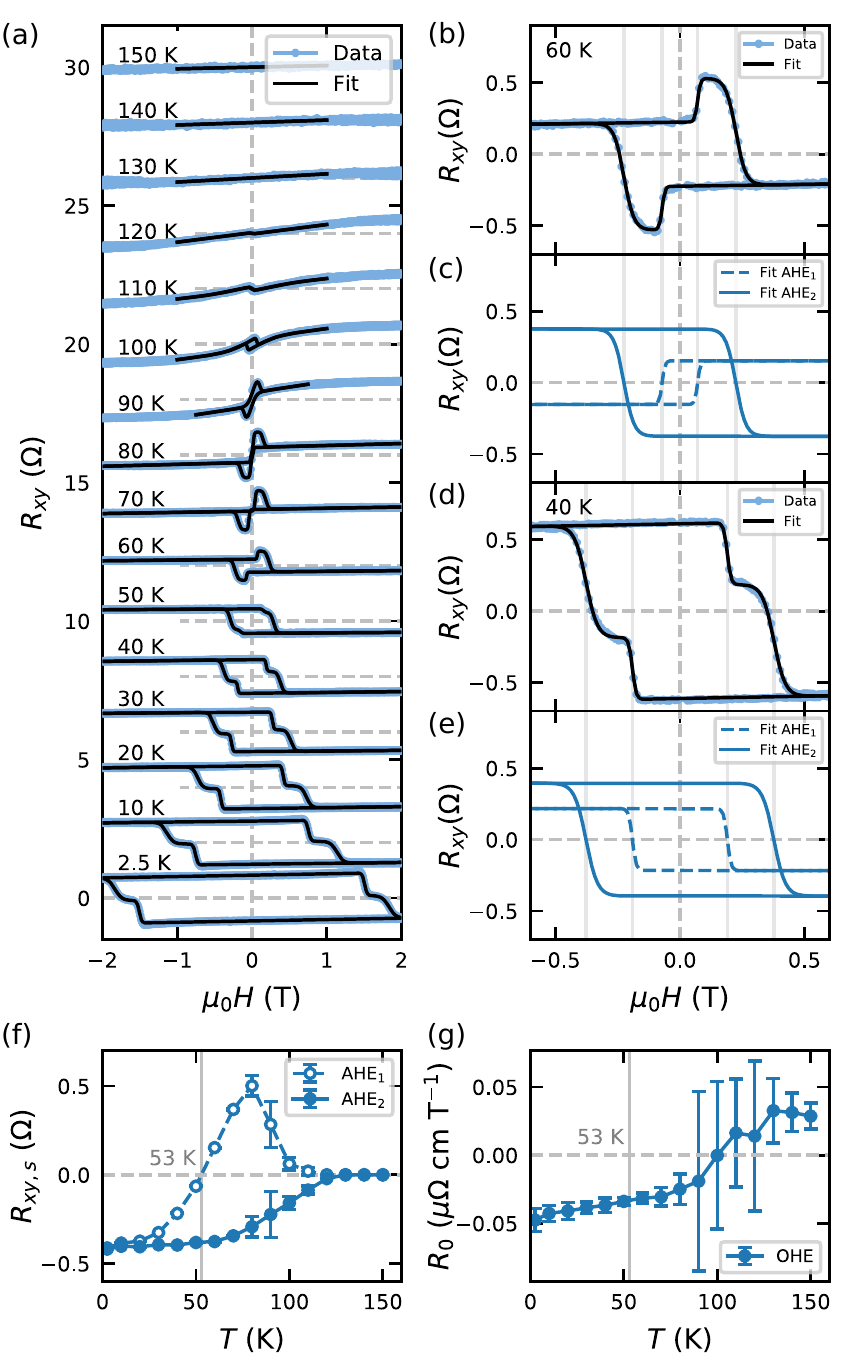}
\centering
\caption{Transport measurements of a nominally 4.0 UC thick SrRuO\textsubscript{3} film. (a) $T-$dependence of $R_{xy}(H)$, showing a transition from the usual `peak' shape below $T_C$, to a `step' shape at 50~K and below. (b) $R_{xy}(H)$ at 60~K showing the usual `peak' shape, with the fitting shown in black. (c) The two components of the fit at 60~K: one $+$ve and one $-$ve AHE. (d) $R_{xy}(H)$ at 40~K showing the `step' shape, with the fitting shown in black. (e) The two components of the fit at 40~K: two $-$ve AHEs. (f) $T-$dependence of the two AHE components. One component experiences a sign change near 53~K, where the anomalous Hall signal changes from step-like at low $T$ to peak-like at high $T$. (g) $T-$dependence of the OHE, found from a linear fit at 2-3~T, the error is the difference between the OHE estimated by a high field and low field fits. The error becomes very large above 90~K, where it becomes difficult to separate the broad paramagnetic AHE from the OHE. The OHE does not change sign near 53~K.
\newline
\newline
\newline
\newline
}
\label{fig:SNU}
\end{figure}

\medskip

\begin{figure*}[!htbp]
\includegraphics[width=\textwidth]{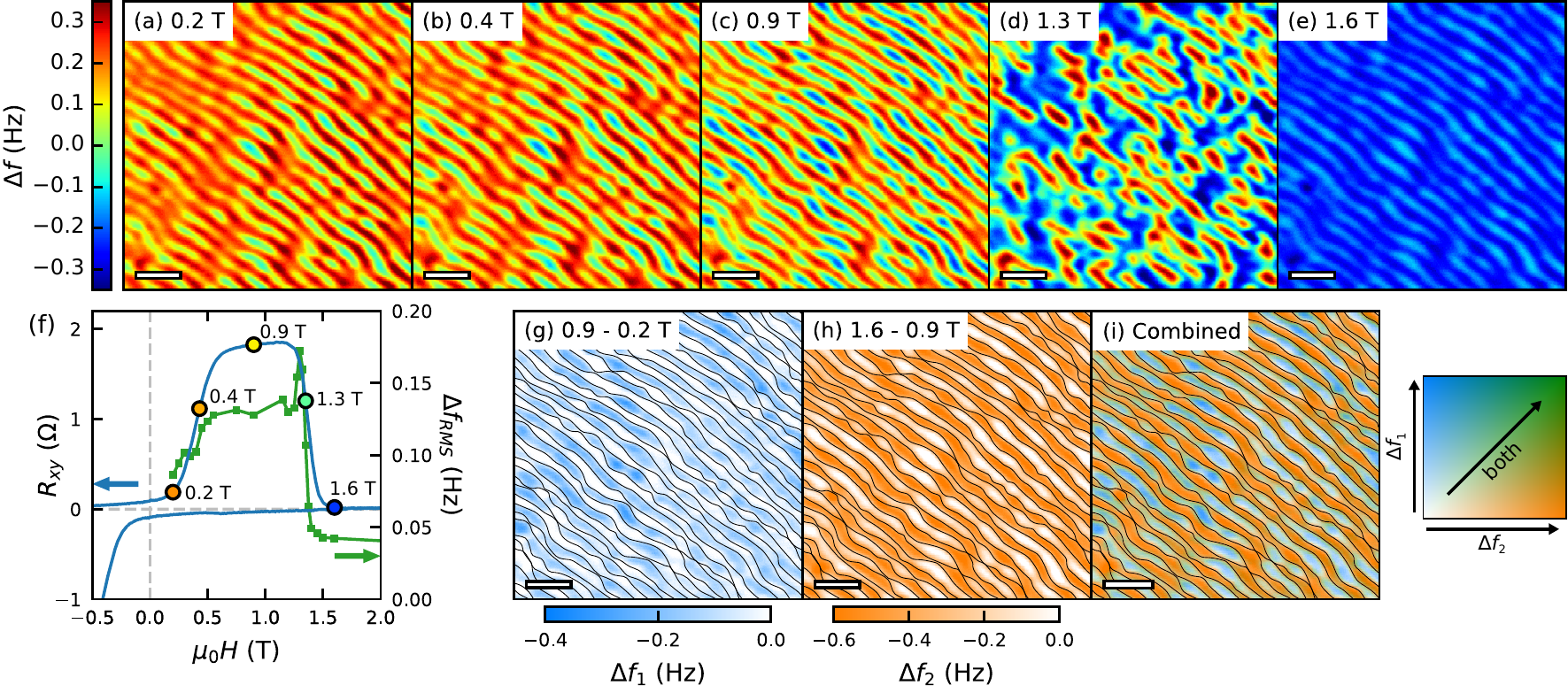}
\centering
\caption{MFM images at 10~K, 0.1~T after applying different fields. Scale bars are 1~$\mu$m. $\Delta f$, the change in resonant frequency, is proportional to out-of-plane stray field gradient. The sign of $\Delta f$ is opposite to the sign of magnetization. Images (a)-(e) show the progression of the magnetic structure of the film through the two transitions. Red/yellow is negatively magnetized, blue/cyan is positively magnetized. (f) shows the anomalous Hall resistance (blue line) of the film as a function of field, with points labeled where the presented MFM images were measured. The RMS deviation of the MFM signal (green squares) shows a peak in inhomogeneity corresponding to the peak in the Hall effect. (g) The change in MFM signal through the first transition, showing only one set of stripes switching. Black lines are added as a guide to the eye. (h) The change in MFM signal through the second transition, showing the other set of stripes switching. The same black lines are a guide to the eye. (i) The images from (g) and (h) combined additively. The MFM signal either changes in the first transition or the second rather than both, meaning there are two spatially magnetic regions with two different coercive fields.
}
\label{fig:MFM1}
\end{figure*}

MFM images show stripe domains in the SrRuO\textsubscript{3} films consisting of two regions with different $H_c$ and magnetic strengths. Figs.~\ref{fig:MFM1}(a)-\ref{fig:MFM1}(e) are MFM images at 10~K, 0.1~T after saturating at $-$ve field and sequentially applying different fields, which are also shown on the $R_{xy}(H)$ loop in Fig.~\ref{fig:MFM1}(f). Following the increasing field: at 0.2~T the film is still negatively magnetized, stripe contrast is present as thinner regions have a lower $M_s$; 0.4~T, weaker regions (yellow-green) are switching; 0.9~T, weaker regions are now positively magnetized (light blue), this switch corresponds to $H_{c1}$ for the $+$ve AHE component (AHE\textsubscript{1}); 1.3~T, stronger magnetic regions (red) begin to switch (blue); 1.6~T, all strong regions have switched to $+$ve magnetization (blue), this corresponds to $H_{c2}$ for the $-$ve AHE component (AHE\textsubscript{2}). The stripe contrast remains even at high field due to the difference in $M_s$ between the two regions.

\medskip

The root mean square (RMS) deviation of the MFM signal is also plotted in Fig.~\ref{fig:MFM1}(f) (green squares). This quantifies the magnetic inhomogeneity in the film - this peak in RMS signal matches well with the peak in the Hall effect.

\medskip

Fig.~\ref{fig:MFM1}(g) shows the change in MFM signal through the first switch, and Fig.~\ref{fig:MFM1}(h) through the second switch - each switch corresponds to one set of stripes changing magnetization. The combination of these two images is shown in Fig.~\ref{fig:MFM1}(i), the lack of overlap demonstrating that the two switches correspond to two spatially separated magnetic regions.

\medskip

These data are consistent with the picture that one UC thickness variations across terrace steps create two magnetic regions with different $M_s$, which results in two AHE channels with different signs and, due to their differing $H_c$, a peak appears in the Hall effect. 

\bigskip
\bigskip
\bigskip
\newpage

\section{Conclusion}

The anomalous Hall coefficient in SrRuO\textsubscript{3} depends strongly on the band structure and magnetization, and can switch sign with parameters that affect these, such as temperature, film thickness, or disorder. In a SrRuO\textsubscript{3} thin film, if there are two magnetic regions with different coercive fields and signs of the AHE, then peaks will appear in Hall effect measurements. Here, 4 to 5 UC thick SrRuO\textsubscript{3} films show peaks in the Hall effect, similar peaks have been observed previously and are sometimes attributed to a THE caused by magnetic skyrmions. SrRuO\textsubscript{3} films in this work show two spatially separated magnetic regions with different $T_C$, $M_s$ and $H_c$ values. The stripe pattern of these two regions indicate they likely result from step-flow growth giving single unit cell thickness variations in the film across terrace steps. These two magnetic regions can explain the peaks in the Hall effect by the superposition of two AHEs with different signs. Additionally, a film showed a smooth transition from peak-like to step-like Hall effect, which is not easily explained by a THE.

These data do not exclude the existence of a THE, however we show that the observation of a peak in the Hall effect does not give unambiguous evidence for a THE caused by magnetic skyrmions, particularly in the case where the magnetic structure of the SrRuO\textsubscript{3} film is inhomogenously modified.

\bigskip

\begin{acknowledgments}
This work is supported by the EPSRC through the Core-to-Core International Network ``Oxide Superspin''  (EP/P026311/1) and the Doctoral Training Partnership Grant (EP/N509620/1). Additional support from the Office of Basic Energy Sciences Division of Materials Sciences and Engineering, US Department of Energy under Award numbers DE-SC0018153, and the Research Center Program of IBS (Institute for Basic Science) in Korea (IBS-R009-D1).
\end{acknowledgments}


\bibliography{zotero_references}

\end{document}


\onecolumngrid

\title{Two-channel anomalous Hall effect in  SrRuO\textsubscript{3}\\Supplementary Information}

\author{Graham Kimbell}
\affiliation{Department of Materials Science \& Metallurgy, University of Cambridge, CB3 0FS, United Kingdom}
\author{Paul M. Sass}
\affiliation{Department of Physics \& Astronomy, Rutgers University, Piscataway, NJ 08854, USA}
\author{Bart Woltjes}
\affiliation{Department of Materials Science \& Metallurgy, University of Cambridge, CB3 0FS, United Kingdom}
\author{Eun Kyo Ko}
\affiliation{Center for Correlated Electron Systems, Institute for Basic Science (IBS), Seoul 08826, Republic of Korea}
\affiliation{Department of Physics \& Astronomy, Seoul National University, Seoul 08826, Republic of Korea}
\author{Tae Won Noh}
\affiliation{Center for Correlated Electron Systems, Institute for Basic Science (IBS), Seoul 08826, Republic of Korea}
\affiliation{Department of Physics \& Astronomy, Seoul National University, Seoul 08826, Republic of Korea}
\author{Weida Wu}
\affiliation{Department of Physics \& Astronomy, Rutgers University, Piscataway, NJ 08854, USA}
\author{Jason W. A. Robinson}
\email[]{jjr33@cam.ac.uk}
\affiliation{Department of Materials Science \& Metallurgy, University of Cambridge, CB3 0FS, United Kingdom}

\date{\today}

\maketitle

\onecolumngrid

\beginsupplement

\FloatBarrier
\section{\SRO growth}

Certain SrRuO\textsubscript{3} films were grown at the Institute for Basic Science in Seoul National University (Fig.~5 of main paper, films labelled as 4.0 or 5.0~UC). A similar growth process is used as growth in the Department of Materials Science and Metallurgy, University of Cambridge, as described in the Methods section, but the SrTiO\textsubscript{3} substrates are held at 700$^\circ$C during deposition, a stoichiometric SrRuO\textsubscript{3} target is used, in-situ reflection high-energy electron diffraction (RHEED) is used to monitor film thickness, and the sample is cooled at $-30^\circ$C~min$^{-1}$ in 13~Pa O\textsubscript{2} after deposition.

Thickness estimates based on both RHEED and growth rate calibration give an estimated $\pm 0.2$~UC uncertainty in thickness. Thickness values quoted in the paper and supplementary information are the nominal values.

\FloatBarrier

\section{Atomic force microscopy}

\begin{figure}[!htbp]
\centering
\includegraphics[width=\textwidth]{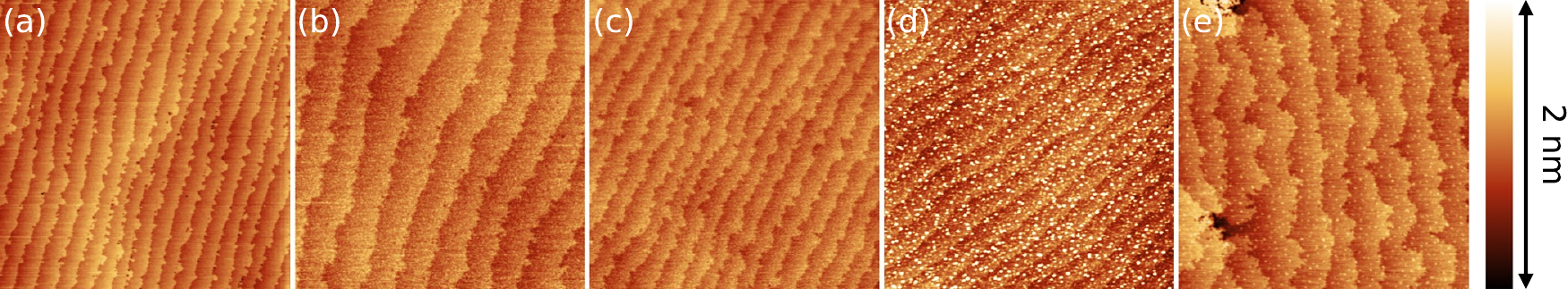}
\caption{Atomic force microscopy scans of (a) 4.0~UC, (b) 4.3~UC, (c) 4.5~UC, (d) 5.0~UC, (e) 27~UC (10~nm) SrRuO\textsubscript{3} films, all scale bars are 1~$\mu$m. Films show a step-and-terrace structure typical of step flow growth. There is typically some instability of the step edges, which will lead to 1~UC variations in sample thickness even in nominally integer UC thick samples. Root-mean-square roughness is below 0.2~nm for all scans, except (d) where some surface debris gives 0.5~nm RMS.  (a), (c), (d) and (e) are the same 4.0, 4.5 and 5.0 UC, and 10~nm samples investigated in the paper. (b) is a twin sample of that measured by MFM in Fig.~6 and Fig.~\ref{fig:supp:MFM1}.}
\label{fig:supplementary-afm}
\end{figure}

\clearpage

\FloatBarrier
\section{X-Ray diffraction and reflection}

\begin{figure}[htbp]
\centering
\includegraphics[width=\textwidth]{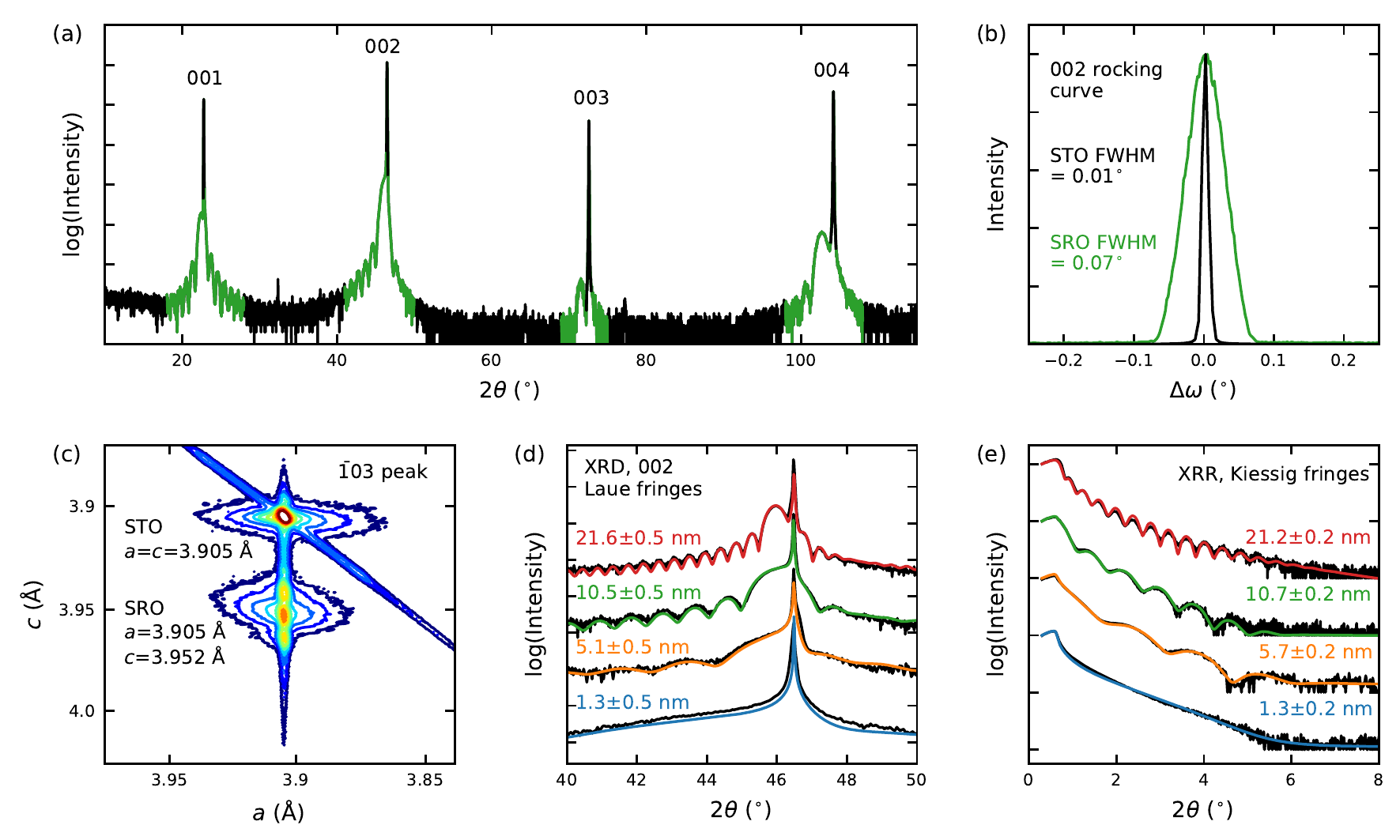}
\caption{X-ray diffraction and reflection of SrRuO\textsubscript{3} films. All intensity scales are arbitrary. (a) 2$\theta$-$\omega$ scan of the 10~nm thick SrRuO\textsubscript{3} film. Film peaks are indicated in green. (b) Normalised rockings curve of the (002) peak of the 10~nm thick film and substrate. (c) Reciprocal space map of the $(\bar{1}03)$ peak of a 10~nm film, the film is fully strained in-plane as evidenced by the same $a$-axis lattice parameter. The $c$-axis parameter is larger than bulk due to Poisson strain. The diagonal streak is caused by detector saturation at the substrate peak. (d) 2$\theta$-$\omega$ scan of the (002) peak of various thickness of SrRuO\textsubscript{3}. The data is shown in black, and dynamical simulation fit shown as coloured lines.  (e) Low angle XRR scans of various thicknesses of SrRuO\textsubscript{3}. The data is in black and simulation fit shown as coloured lines.}
\label{fig:xray}
\end{figure}

The structural properties of SrRuO\textsubscript{3} films are investigated using X-ray diffraction (XRD, Panalytical Empyrean X-ray diffractometer with primary 2-bounce monochromator) and X-ray reflection (XRR, Bruker D8 diffractometer). A 2$\theta-\omega$ scan of a 10-nm-thick sample is shown in Fig.~\ref{fig:xray}(a), only the $(00l)$-type peaks are detected, indicating the film has grown epitaxially on the $(001)$-oriented STO substrate. The film is well-aligned to the substrate - a rocking curve on the $(002)$ peak (Fig.~\ref{fig:xray}(b)) gives a full-width-half-maximum (FWHM) of 0.07$^{\circ}$; this is an upper limit on the mosaic spread in the film. A reciprocal space map of the $(\bar{1}03)$-peak (Fig.~\ref{fig:xray}(c)) gives both in-plane and out-of-plane lattice parameters of the film and substrate. The in-plane lattice parameter of the film is the same as the substrate (3.905~\AA), so the film is fully strained to the substrate. Poisson strain causes the $c$-axis parameter ($\sim 3.95$~\AA) to deviate from its bulk value ($\sim 3.93$~\AA). Film thicknesses are determined by fitting full dynamical simulations to high-angle XRD peaks using Leptos software (Fig.~\ref{fig:xray}(d)), and low angle XRR fringes using GenX software \cite{bjorck_genx_2007} (Fig.~\ref{fig:xray}(e)). These figures show the difficulty in determining the thickness and lattice parameter of very thin ($<5$~nm) films as peaks and fringes become broad and weak, hence extrapolation from thicker films is necessary.

\clearpage
\FloatBarrier
\section{Transport measurements}

\FloatBarrier
\subsection{Longitudinal resistivity}

\begin{figure}[!htbp]
\centering
\includegraphics[]{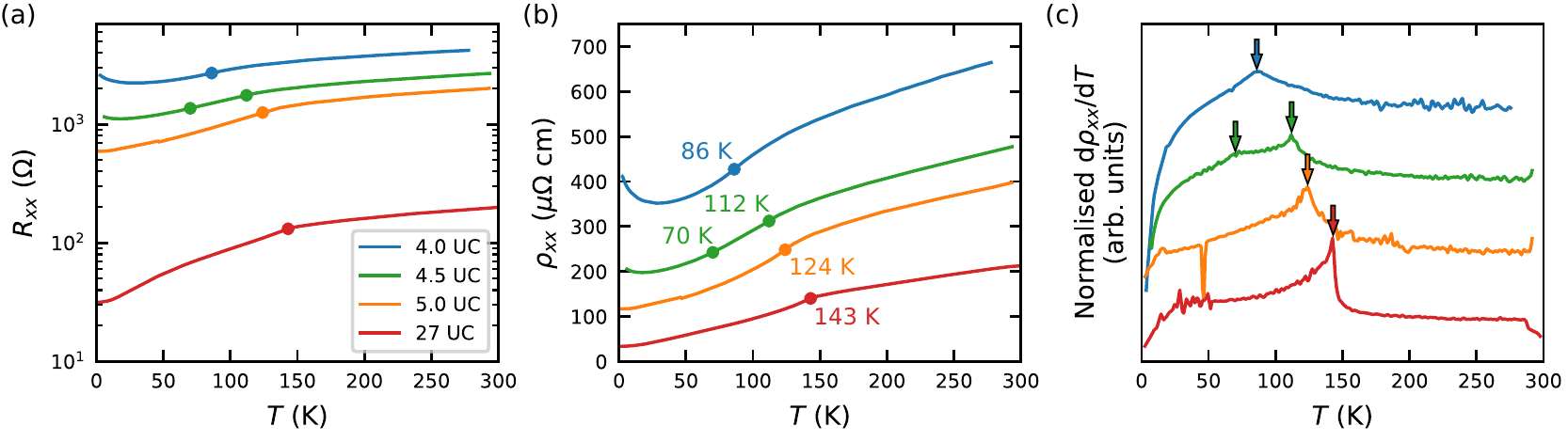}
\caption{(a) $R_{xx}(T)$, (b) $\rho_{xx}(T)$ and (c) d$\rho_{xx}(T)$/d$T$ for SrRuO\textsubscript{3} samples. The derivative in (c) is normalised and curves are shifted vertically for clarity. The is a kink in the resistivity at the Curie temperature ($T_C$) of the film, indicated by filled circles in (a) and (b), and arrows in (c). As film thickness decreases, resistivity increases and $T_C$ decreases and broadens. Only one $T_C$ is clearly visible here for the nominally 4.0~UC film, whereas magnetic measurements revealed multiple transitions (Fig.~\ref{fig:SNUmag}). Two transitions are visible in the 4.5~UC film, likely from separate 4~UC and 5~UC regions.}
\label{fig:RT-all}
\end{figure}

\FloatBarrier
\subsection{Antisymmetrisation}

\begin{figure}[!htbp]
\includegraphics[]{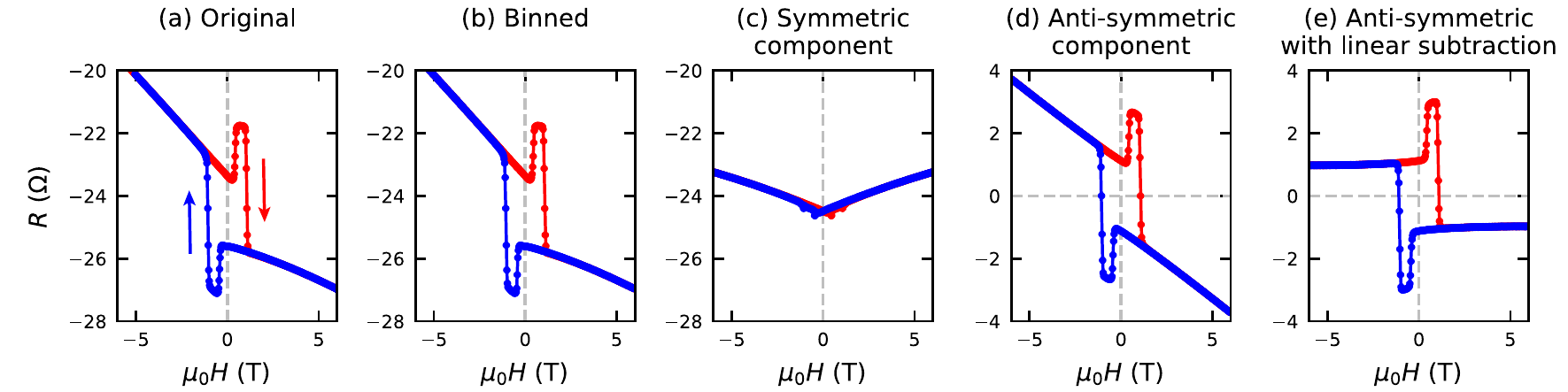}
\centering
\caption{(a) Raw measured data of 4.5~UC film at 10~K. (b) Binned data with 400 bins over 6~T range. (c) Symmetric component of Hall effect data, this is assumed to be due to a longitudinal component of resistance and is ignored. (d) Anti-symmetric component of Hall data, this is taken to be the true Hall effect. (e) Anti-symmetric component with a high field (5-6~T) linear fit subtracted. This linear part is assumed to be the OHE, and the remaining Hall effect is assumed to be due to the AHE.}
\label{fig:antisym}
\end{figure}

Hall measurements are made in a Van der Pauw geometry, and some longitudinal resistance component remains in the measurement. In order to extract the Hall component the data is first binned (since data in the positive and negative sweeps are not measured at exactly the same field values) then split into the symmetric and anti-symmetric component by:
\begin{equation}
    R_{sym}(H) = \left(R(H) + R(-H)\right)/2
\end{equation}
\begin{equation}
    R_{asym}(H) = \left(R(H) - R(-H)\right)/2
\end{equation}
In all figures in the paper, the high-field linear part of the anti-symmetric component, assumed to be caused by the OHE, is subtracted by a linear fit in high field region (usually 5-6~T). An example of this antisymmetrisation and linear subtraction process is shown in Fig.~\ref{fig:antisym}.

\FloatBarrier
\subsection{Additional anomalous Hall effect measurements}

\begin{figure}[htbp]
\includegraphics{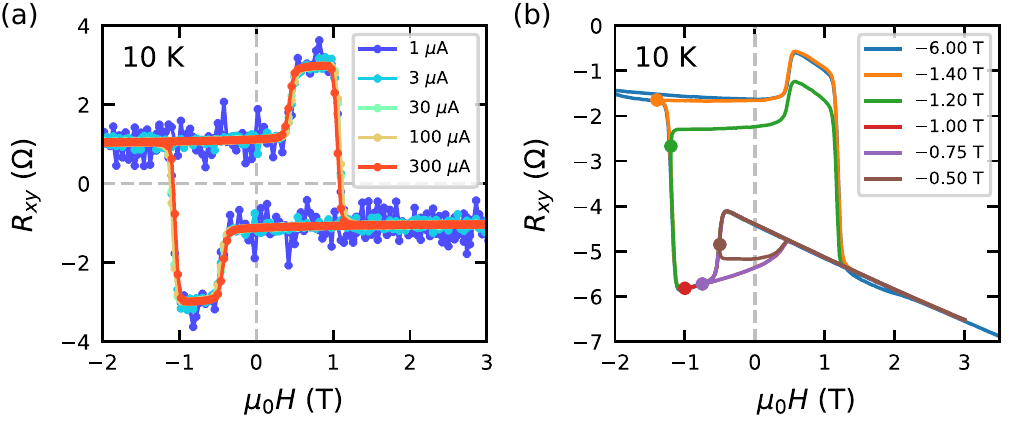}
\centering
\caption{(a) $R_{xy}(H)$ loops of a 4.5~UC film at 10~K using different applied longitudinal currents. There is no clear change in the shape of the loop, only in the signal-to-noise ratio. (b) Major and minor $R_{xy}(H)$ loops of at 10~K of a different sample which also showed peaks in the Hall effect. The data is raw so includes longitudinal magnetoresistance component. The circle marker and legend indicates the minimum field reached before reversing the magnetic field in each minor loop.}
\label{fig:currentdep}
\end{figure}

Current-dependent $R_{xy}(H)$ loops have been observed for ultrathin SrRuO\textsubscript{3} in which the `THE' peak decreases with increasing current density, whilst the AHE component remains constant, reported as a manifestation of the skyrmion Hall effect \cite{sohn_emergence_2018}. Such current dependence is not observed here, as shown in Fig.~\ref{fig:currentdep}(a), despite using a comparable range of currents. Samples are $5\times 5$~mm, and current is applied across the diagonal of the film in a Van der Pauw geometry.

Fig.~\ref{fig:currentdep}(b) shows raw Hall effect minor loops, with the behavior expected for two uncoupled magnetic regions, however we note that similar minor loops have been previously explained in terms of nucleation and annihilation of skyrmions \cite{ludbrook_nucleation_2017}.

\FloatBarrier
\section{Magnetism}
\FloatBarrier
\subsection{Corrections for volumetric magnetic data}
\FloatBarrier

\begin{figure}[htbp]
\centering
\includegraphics{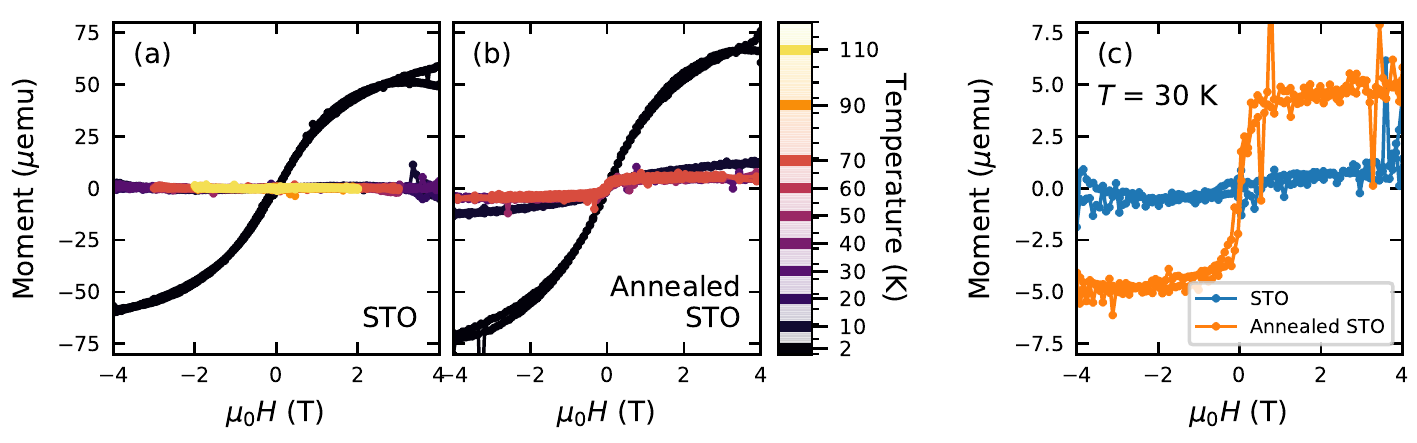}
\caption{(a) Control $M(H)$ measurements of an as-bought SrTiO\textsubscript{3} (STO) substrate, and a substrate which has undergone the same cleaning and heating process as the SrRuO\textsubscript{3} films. Both show very large paramagnetic component below 10~K. In the cleaned and annealed film, there is a small ferromagnetic component which has no observable temperature dependence between 2-70~K. (b) Comparison of as-bought and cleaned and annealed film at 30~K. There is a $\sim 5$~$\mu$emu ferromagnetic component of the STO after cleaning and annealing.}
\label{fig:mpmscorr2}
\end{figure}

Volume magnetometry measurements were made in a Quantum Design MPMS3 system. In order to separate background and film components of magnetism, $M(H)$ loops of SrTiO\textsubscript{3} substrates were measured in the as-bought state (Fig. \ref{fig:mpmscorr2}(a)), as well as after undergoing the same cleaning and deposition conditions as the films (Fig. \ref{fig:mpmscorr2}(b)). In both cases, the SrTiO\textsubscript{3} displayed a large paramagnetic component below 10~K, which can be approximated by a broad Langevin function. SrTiO\textsubscript{3} also exhibits a small ferromagnetic-like magnetic moment after cleaning and annealing, a comparison at 30~K (above the temperature where paramagnetism dominates) is shown in Fig.~\ref{fig:mpmscorr2}(c). This is approximated by a temperature-independent sharp Langevin function. Both the broad low temperature paramagnetic component and a ferromagnetic component which depends on solvent cleaning (with $T_C >$ room temperature) has been observed in literature \cite{khalid_ubiquity_2010}.

\begin{figure}[htbp]
\centering
\includegraphics{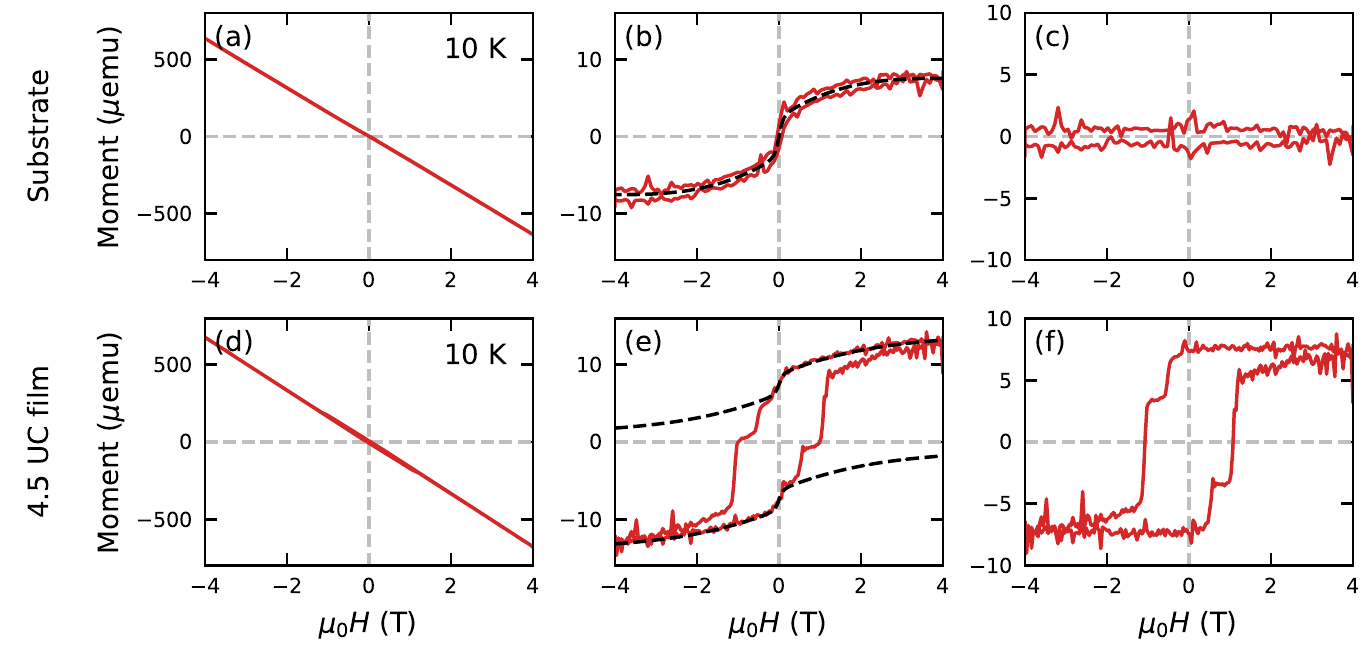}
\caption{(a) Raw $M(H)$ data of control SrTiO\textsubscript{3} substrate. (b) Raw $M(H)$ data of a 4.5~UC SrRuO\textsubscript{3} sample. (c) Substrate $M(H)$ with a linear subtraction. There is still a ferromagnetic component, fit by the summation of two langevin functions (one broad, one sharp) shown by the black dotted line. (d) Film $M(H)$ with linear subtraction, the dotted black shows the two langevin functions assumed to be the substrate component of the ferromagnetism, which is subtracted from the data. (e) Substrate data with linear + langevin(broad) + langevin(sharp) removed. (f) Film data with the assumed substrate component removed. The remaining moment is attributed to the ferromagnetic SrRuO\textsubscript{3} films.}
\label{fig:mpmscorr3}
\end{figure}

The process of removing the substrate component of ferromagnetism from film measurements is shown in Fig.~\ref{fig:mpmscorr3}. The low field Langevin component of the substrate was assumed to be temperature independent, and the same subtraction is used for every $M(H)$ loop. A subtraction is made for the broad Langevin at 2 and 10~K measurements (included in Fig. \ref{fig:mpmscorr3}); it is considered insignificant at $\geq 20$~K.

\clearpage

\FloatBarrier
\subsection{Volume magnetometry}
\FloatBarrier

\begin{figure}[htbp]
\centering
\includegraphics{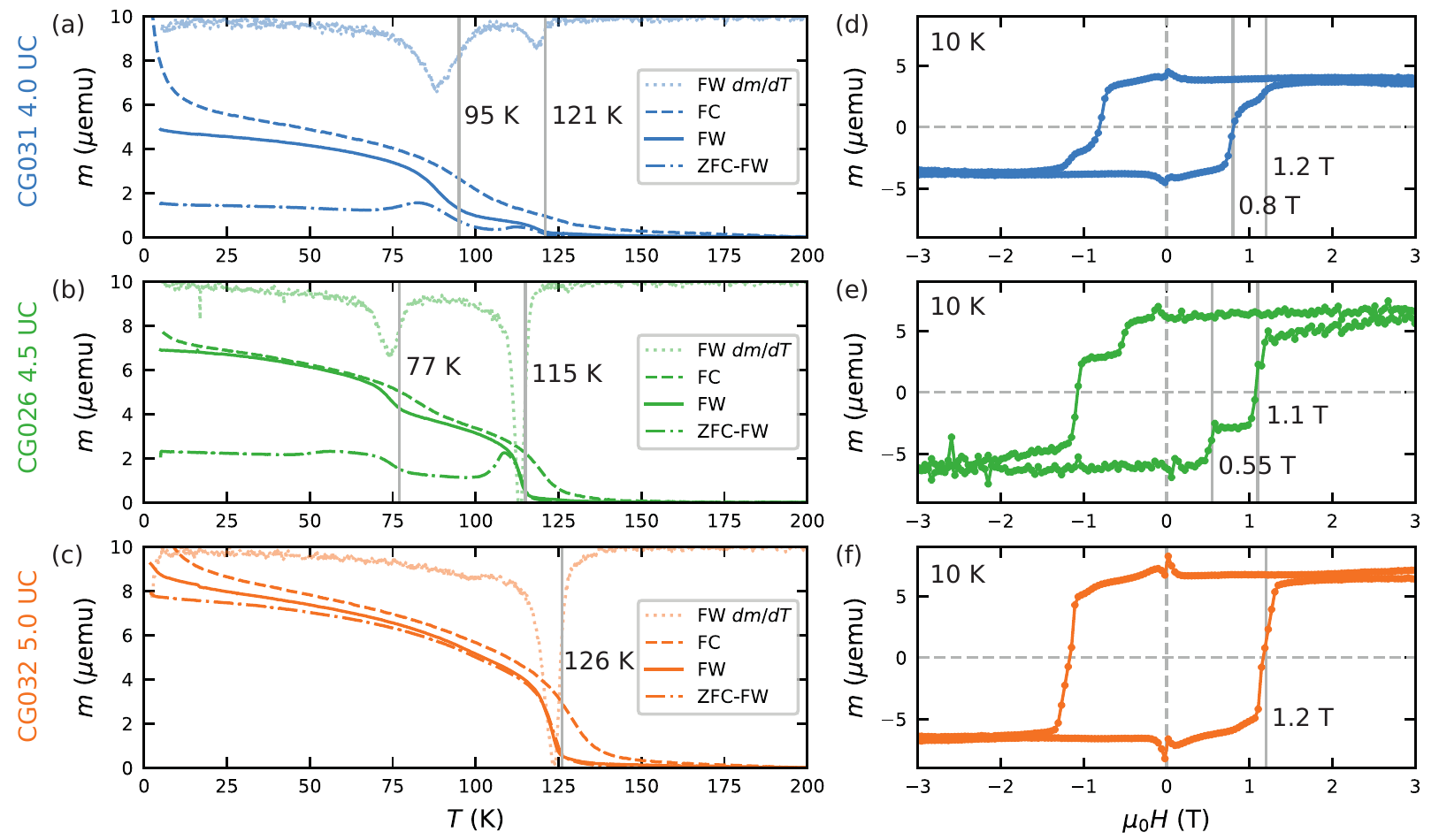}
\caption{(a)-(c) Moment versus $T$ for 4.0 (a), 4.5 (b) and 5.0 (c) UC thick films. FC is field cooling in 100~mT, FW is field warming in 5~mT after saturating at 7~T, ZFC-FW is field warming in 5~mT after cooling in zero field. Curves are fixed at $m(200) = 0$. FW $dm/dT$ is the derivative of the FW curve with arbitrary units and shifted vertically for clarity. (a) and (b) show two transitions, likely corresponding to 4 UC (lower $T_C$) and 5 UC (higher $T_C$) regions. (d)-(f) Moment versus $H$ at 10~K for 4.0 (a), 4.5 (b) and 5.0 (c) UC thick films. (a) and (b) show two coercive fields ($H_c$) due to 4 UC (low $H_c$) and 5 UC (high $H_c$) regions. Only one $H_c$ is clear in (c). Low field artifacts are from substrate subtraction (see Fig. \ref{fig:mpmscorr3}).}
\label{fig:SNUmag}
\end{figure}

\begin{figure}[htbp]
\centering
\includegraphics{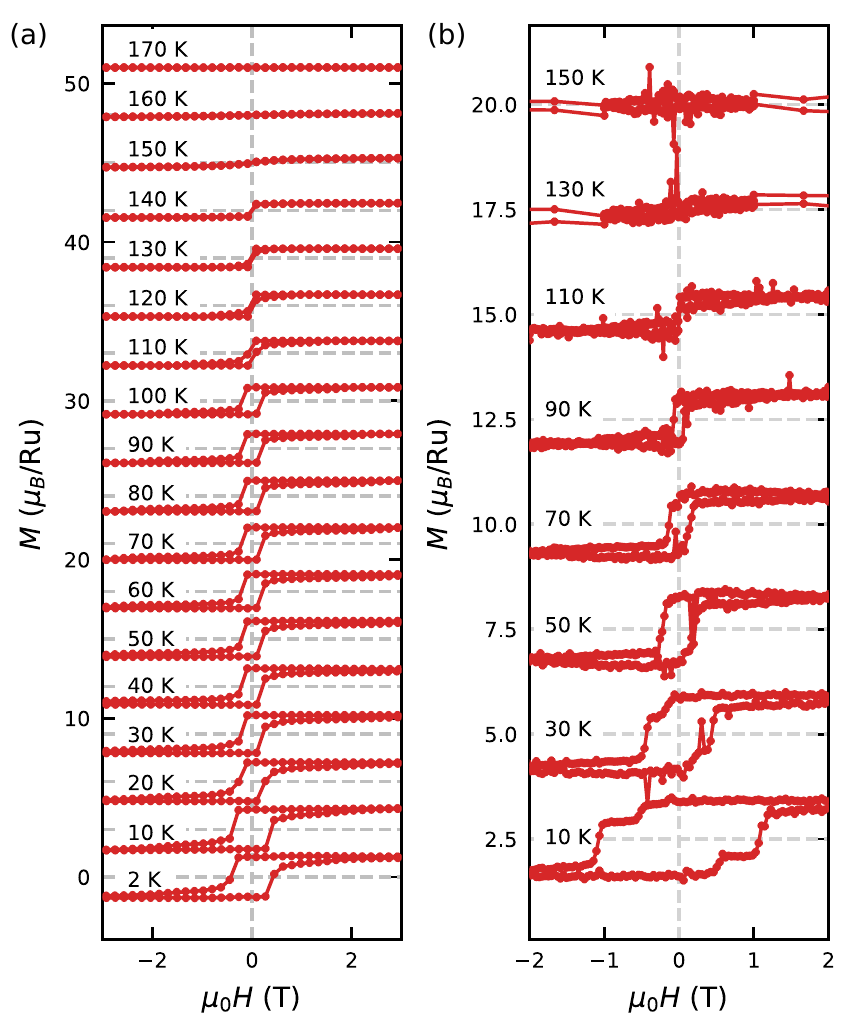}
\caption{(a) $M(H)$ at different $T$ for a 10 nm sample, used to calculate $M_s$ in Fig. 3(c) of the main paper. (b) $M(H)$ at different $T$ for a 4.5 UC sample, the 10~K loop is shown in Fig. 4(d) of the main paper. The two magnetic switches are only clear at low temperature ($<30$~K) due to the low signal:noise for volume magnetometry on very weakly magnetic films.}
\label{fig:MHTs}
\end{figure}

\clearpage

\FloatBarrier
\subsection{Magnetic Force Microscopy}
\FloatBarrier

\begin{figure}[htbp]
\centering
\includegraphics[width=\textwidth]{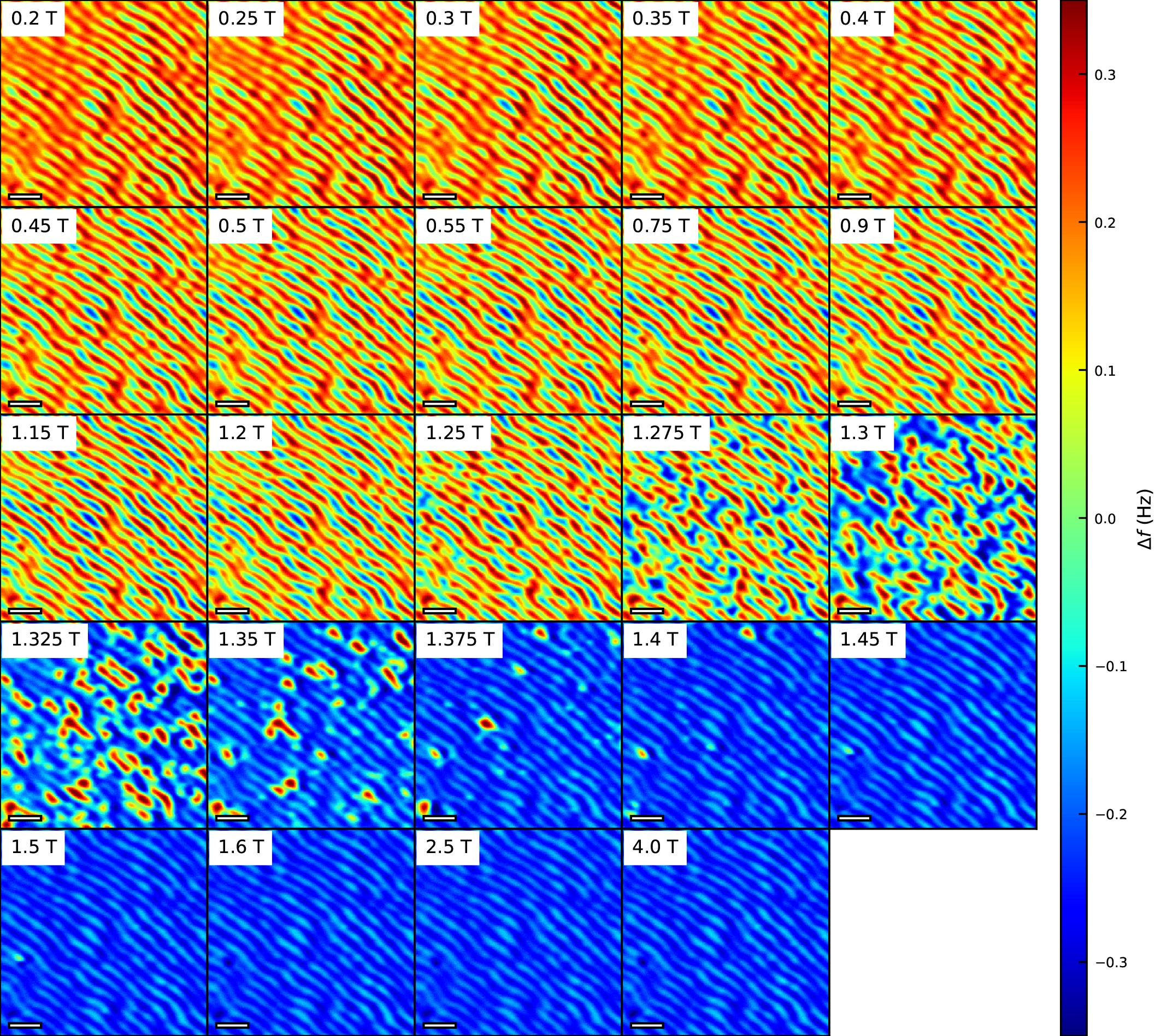}
\caption{MFM images as a function of field of a nominally 4.3 UC sample. The scale bars are 1~$\mu$m. The sign of $\Delta f$ is opposite to the sign of magnetization.}
\label{fig:supp:MFM1}
\end{figure}

\begin{figure}[htbp]
\centering
\includegraphics[width=\textwidth]{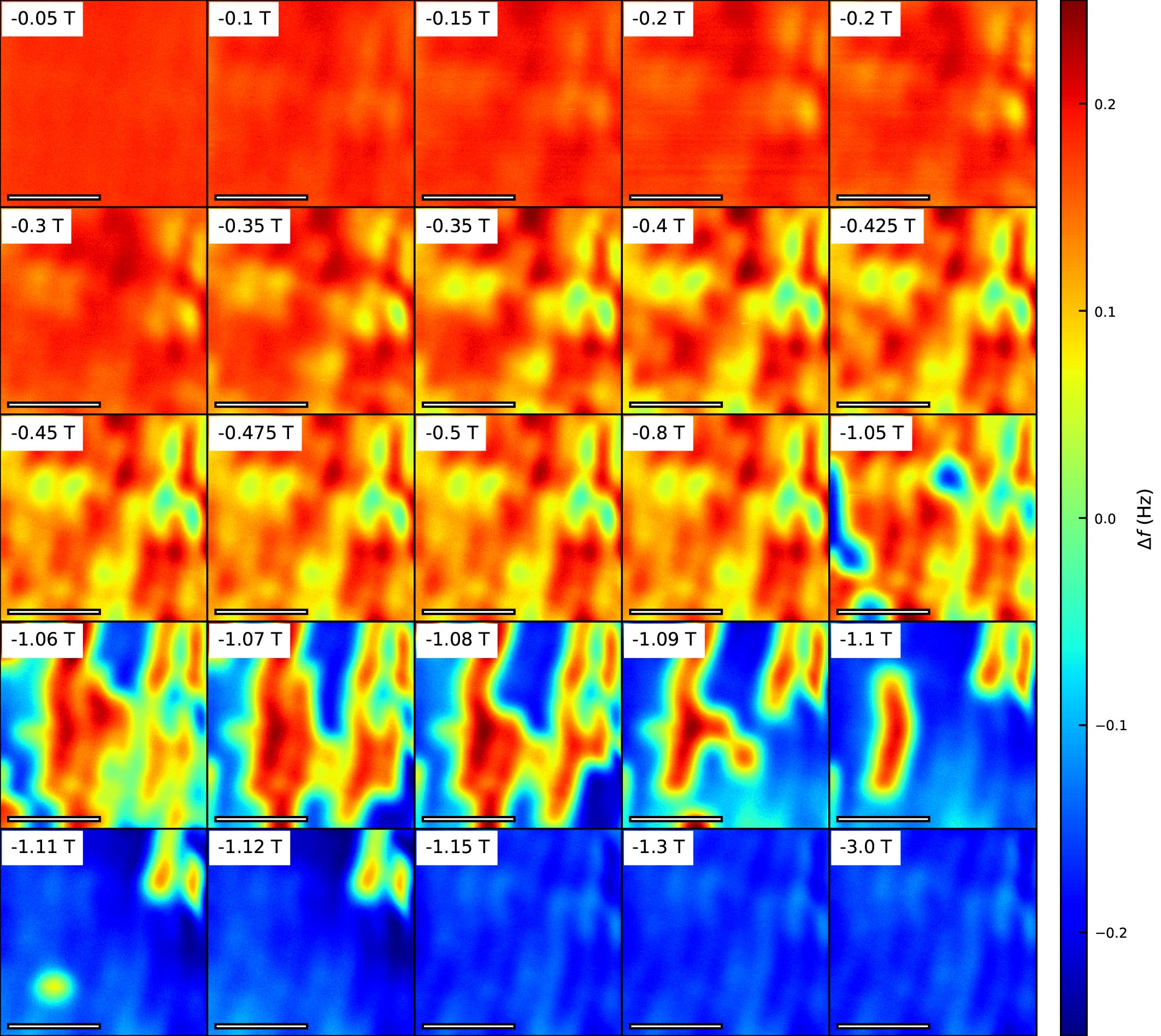}
\caption{MFM images as a function of field of a nominally 4.5 UC sample. The scale bars are 1~$\mu$m, and in this case the scale had to be approximated from AFM measurements. Here, the sign of $\Delta f$ is the same as the sign of magnetization (the difference between this and Fig.~\ref{fig:supp:MFM1} is due to the cantilever tip being negatively magnetized).}
\label{fig:supp:MFM2}
\end{figure}

This section shows full results of MFM measurements. Fig.~\ref{fig:supp:MFM1} shows the complete set of results from a nominally 4.3~UC sample (used in Fig.~6 of the main paper), and Fig.~\ref{fig:supp:MFM2} shows a set of measurements from a 4.5~UC sample (used in Fig.~4 of the main paper). In both cases there is a stripe contrast, where one region is magnetically weaker and switches at a lower field, and the other is magnetically stronger and switches at a higher field. The stripe contrast persists even in the high field saturated state because the two regions have different saturation magnetizations.

\clearpage
\FloatBarrier
\section{Effects of age}
\FloatBarrier

\begin{figure}[htbp]
\centering
\includegraphics[]{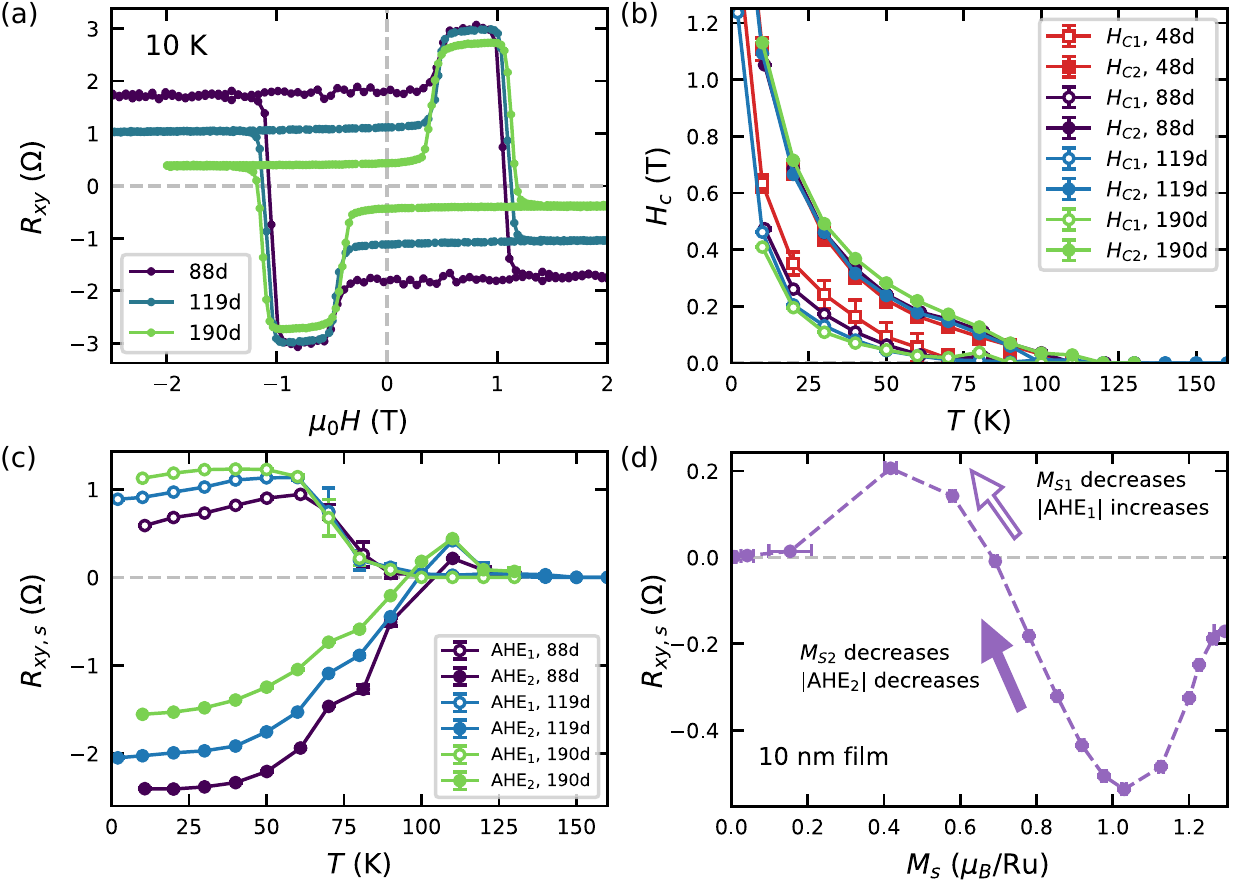}
\caption{Changes in the film over time, $x$~d indicates the number of days after growth of the film. (a) Change in $R_{xy}(H)$ over time in the same film measured at 10~K. (b) $T-$dependence of the two coercive fields over time in the same film, as measured in magnetometry (red squares) and transport measurements (circles). (c) $T-$dependence of the spontaneous AHE over time. (d) Spontaneous AHE versus $M_s$ for a thicker film. If the two AHE channels in the thinner film decrease in $M_s$ over time, then the smaller $M_s$ channel can become more negative, and the larger $M_s$ channel can become less positive, leading to the observed behavior.}
\label{fig:degredation}
\end{figure}

Over the course of measurements it appeared that the coercive fields and anomalous Hall signals from the films changed over time. The lower coercive field decreased over time, and upper coercive field increased over time, an example is shown in Fig.~\ref{fig:degredation}(a), and the full temperature dependence including magnetometry data is shown in Fig. \ref{fig:degredation}(b), though this may also be related to differences in field sweep rates \cite{kan_field-sweep-rate_2020}. Furthermore, from Fig. \ref{fig:degredation}(a) it appears as though the saturated AHE decreases in size, whereas the Hall effect peak size increases by a similar amount. In Fig.~\ref{fig:degredation}(c) the Hall effect is fitted with the two AHE model (component 1, lower $M_s$, $+$ve AHE; component 2, larger $M_s$, $+$ve then $-$ve AHE), it appears that AHE\textsubscript{1} increases in magnitude, AHE\textsubscript{2} decreases in magnitude, and the temperature crossover decreases over time. This is consistent with both channels decreasing in magnetization over time as illustrated in Fig.~\ref{fig:degredation}(d). A direct conversion between magnetization and Hall effect difficult as it requires precise measurements of the area fraction and resisitivity of each region. This degradation of magnetization is perhaps due to the loss of oxygen, or damage from various heating/cooling cycles, cleaning in solvents, and exposure to atmosphere.

\FloatBarrier

\bibliographystyle{unsrt}
\bibliography{zotero_references}